\documentclass{aipproc}
\usepackage{amsmath,amssymb}
\usepackage{amsfonts}
\layoutstyle{8x11double}

\begin{document}
\title{Reconstructing SUSY and R-Neutrino Masses in SO(10)}
\classification{12.10.Dm, 12.60.Jv, 14.60.Pq}
\keywords      {Supersymmetry, Grand Unified Theories, Neutrinos}

\author{F. Deppisch}{
address={School of Physics and Astronomy, University of Manchester, Manchester M13 9PL, UK}
}
\author{A. Freitas}{
address={Department of Physics \& Astronomy, University of Pittsburgh, PA 15260, USA},
altaddress = {HEP Division, Argonne National Laboratory, Argonne, IL 60439, USA}
}
\author{W. Porod}{
address={Inst. Theor. Physik und Astrophysik, Universit\"at W\"urzburg, D-97074 W\"urzburg, Germany}
}
\author{P.M. Zerwas}{
address={Inst. Theor. Physik E, RWTH Aachen U, D-52056 Aachen, Germany},
altaddress = {Deutsches Elektronen-Synchrotron DESY, D-22603 Hamburg, Germany}
}

\begin{abstract}
We report on the extrapolation of scalar mass parameters in the lepton sector to reconstruct SO(10) scenarios close to the unification scale. The method is demonstrated for an example in which SO(10) is broken directly to the Standard Model, based on the expected precision from coherent LHC and ILC collider analyses. In addition to the fundamental scalar mass parameters at the unification scale, the mass of the heaviest right-handed neutrino can be estimated in the seesaw scenario.
\end{abstract}

\maketitle

\subsection{Theoretical basis}

A natural explanation of the very light neutrino masses in relation to the electroweak scale is offered by the seesaw mechanism \cite{seesaw}, which naturally suggests the symmetry group SO(10) as the grand unification group \cite{so10}.

In this report, which summarizes the results of Ref.~\cite{Deppisch:2007xu}, we focus on a simple model incorporating one-step symmetry breaking from SO(10) down to the Standard Model SM gauge group, \(\text{SO(10)} \to \text{SM}\), at the GUT scale $\Lambda_{\mathcal U} \approx 2 \cdot 10^{16}$ GeV where the gauge couplings unify in SUSY. We implicitly assume that a number of fundamental problems \cite{Raby} are solved without interfering strongly with the key points of the present analysis, i.e. mechanisms leading to doublet-triplet splitting and suppressing proton decay. Such solutions may include extended Higgs sectors or higher-dimensional Planck-scale suppressed operators.

The SUSY SO(10) model is characterized by the following part of the superpotential which involves matter fields:
\begin{equation}
	{\mathcal W}_{\mathcal{M}} = 
	  Y_{10}  \mathbf{16} \cdot \mathbf{16} \cdot \mathbf{10}
	+ Y_{10'} \mathbf{16} \cdot \mathbf{16} \cdot \mathbf{10'}
	+ Y_{126} \mathbf{16} \cdot \mathbf{16} \cdot \mathbf{126}.  \nonumber
\end{equation}
The matter superfields of the three generations belong to 16-dimensional representations of SO(10). Two Higgs-10 fields generate masses separately for up- and down-type fermions at the electroweak scale. The masses of the heavy R-neutrino superfields are generated by a 126-dimensional Higgs, resulting in a seesaw type I mechanism of light neutrino mass generation.

The superpotential is supplemented by soft SUSY breaking masses for the fermion and Higgs multiplets, assumed to be universal at the unification scale, 
\begin{equation}\label{eq:ScalarUniversality}
	m_{16} = m_{10} = m_{10'} = m_{126} = M_0 \,.
\end{equation}
However, the breaking of the SO(10) symmetry group to the SM group generates GUT D-terms $D_{\mathcal{U}}$ such that the boundary conditions at the GUT scale read \cite{Cheng:1994bi}
\begin{eqnarray}
	m_{L}^2     = M_0^2 - 3 \, D_{\mathcal{U}} \nonumber\\
	m_{E}^2     = M_0^2 + 2 \, D_{\mathcal{U}} \\
	m_{\nu_R}^2 = M_0^2 +5 \, D_{\mathcal{U}}  \nonumber
\end{eqnarray}
for the slepton L-isodoublet and R-isosinglet fields. The D-term is of the order of the soft SUSY breaking masses and we will treat $D_{\mathcal{U}}$ as a free parameter.

\subsection{The neutrino sector}
The heavy right-handed neutrino masses are related to the light neutrino masses by the Yukawa matrix $Y_\nu$ in the seesaw mechanism. Neglecting higher-order effects in the calculation of the Majorana neutrino mass matrix, it follows from the Higgs-10 SO(10) relation
\begin{equation}\label{eqn:YukawaUnification}
	Y_\nu = Y_u
\end{equation}
between the neutrino and up-type quark Yukawa matrices that \( Y_\nu \approx \text{diag}(m_u,m_c,m_t)/v_u \) holds approximately for the neutrino Yukawa matrix at the GUT scale; $v_u = v\sin\beta$, with $v$ and $\tan\beta$ being the familiar vacuum and mixing parameters in the Higgs sector. Quark mixing and RG running effects in the neutrino sector are neglected in the analytical approach but properly taken into account in the numerical analysis.

The effective mass matrix of the light neutrinos is constrained by the results of the neutrino oscillation experiments:
\begin{equation}\label{eqn:mnu}
	m_\nu=
	U_\text{MNS}^* \cdot
	\text{diag}(m_{\nu_1},m_{\nu_2},m_{\nu_3}) \cdot
	U_\text{MNS}^\dagger.
\end{equation}
We will assume the normal hierarchy for the light neutrino masses $m_{\nu_i}$, 
and for the MNS mixing matrix the tri-bimaximal form.
From the seesaw relation
\begin{equation}\label{eqn:SeesawRelation}
	M_{{\nu}_R} = Y_\nu m_\nu^{-1} Y_\nu^T \cdot v^2_u,
\end{equation}
the heavy Majorana R-neutrino mass matrix $M_{{\nu}_R}$ can now be calculated.
\begin{figure}
\begin{minipage}{\columnwidth}
\centering
\includegraphics[clip,width=1\textwidth]{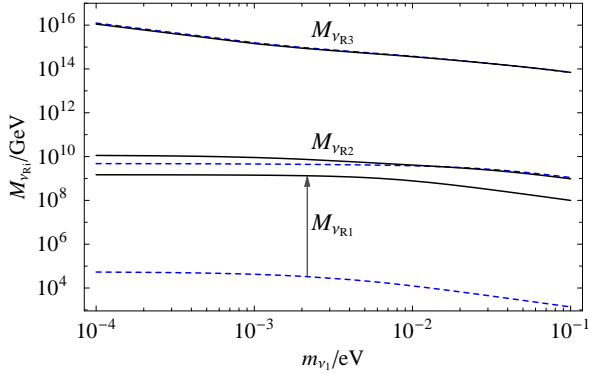}
\end{minipage}
\caption{\it Masses of right-handed neutrinos $M_{\nu_{Ri}}$ as functions of the lightest neutrino mass \(m_{\nu_1}\). The dashed (blue) lines assume perfect Yukawa unification, Eq.~(\ref{eqn:YukawaUnification}). The solid (black) lines indicate shifts of the $\nu_R$ masses if the Yukawa identity Eq.~(\ref{eqn:YukawaUnification}) is violated by an additional contribution $\sim 100\text{ MeV}/v$.}
\label{fig:Mivsm1}
\end{figure}
Solving Eq.~(\ref{eqn:SeesawRelation}) for the eigenvalues $M_{\nu_{Ri}}$ $(i=1,2,3)$, the heavy Majorana masses are determined by the up-quark masses $m_{u,c,t}$ at the GUT scale and the lightest neutrino mass \(m_{\nu_1}\), assuming best fit values for the light neutrino squared mass differences. The mass spectrum of the R-neutrinos is strongly hierarchical in this scenario, \(M_{\nu_{R3}} : M_{\nu_{R2}}  :M_{\nu_{R1}} \sim m_t^2:m_c^2:m_u^2\). The numerical evaluation, including refinements like RG running effects, is displayed in Fig.~\ref{fig:Mivsm1} for a wide range of $m_{\nu_1}$ values. Modifying the relation between the neutrino and up-type quark Yukawa couplings, Eqs.~(\ref{eqn:YukawaUnification}), ad-hoc by a small additional term, $Y_\nu=Y_u+\mathcal{O}(100\text{ MeV})/v$, associated potentially with a more complex Higgs scenario, Planck-scale suppressed contributions or non-perturbative effects, the first generation R-neutrino masses is lifted to ${\mathcal{O}}(10^{10}$ GeV), while the heavier second/third generation masses remain practically unchanged.

\subsection{The scalar sector}
To leading order, the solutions of the RG equations for the masses of the scalar selectrons, can be expressed  in terms of the universal scalar mass $M_0$, the gaugino mass \(M_{1/2}\), and the GUT and electroweak D-terms, $D_{\mathcal{U}}$ and \(D_{EW}=M_Z^2/2 \, \cos2\beta\), respectively\footnote{In the following we omit the discussion of sneutrino and Higgs masses and their role in reconstructing GUT scale parameters, but their effect is included in the numerical results. For further details see Ref.~\cite{Deppisch:2007xu}.},
\begin{eqnarray}\label{eq:RGfirstgen}
  	m_{\tilde e_R}^2  \!\!\! &=&  \!\!\!
	  M_0^2  
	+ \,\,\,\, D_{\mathcal{U}} 
	+ \alpha_R M_{1/2}^2 
	- \tfrac{6}{5}S' 
	- 2s_W^2 D_{EW},                     \\
        m_{\tilde e_L}^2 \!\!\! &=& \!\!\!
	  M_0^2  
	- 3D_{\mathcal{U}} 
	+ \alpha_L M_{1/2}^2 
	+ \tfrac{3}{5}S' 
	- c_{2W} D_{EW}.            
\end{eqnarray}
The coefficients \(\alpha_L\) and \(\alpha_R\) are determined by the gaugino/gauge boson loops in the RG evolution from the SUSY scale $\tilde{M}$ \cite{spa} to the unification scale, and the numerical evaluation yields \(\alpha_R \approx 0.15\) and \(\alpha_L \approx 0.5\). The universal gaugino mass parameter $M_{1/2}$ can be pre-determined in the chargino/neutralino sector. The non-universal initial conditions in the evolution due to the D-terms generate the small generation-indepen\-dent corrections \(S'\), 
{\it cf.} Ref.~\cite{Deppisch:2007xu}.

\begin{table}
\centering
\begin{tabular}{lcc}
\hline
Parameter                       & Ideal  & 1$\sigma$ Error \\
\hline
$M_0$ [GeV]                     &   90.0 & 0.25            \\
$M_{1/2}$ [GeV]                  &  250.0 & 0.4             \\
$A_0$ [GeV]                     & -640.0 & 13.0            \\
$\tan\beta$                     &   10.0 & 1.0             \\
$\sqrt{-D_{\mathcal{U}}}$ [GeV] &   30.0 & 0.9             \\
\hline
\(\Lambda_{\mathcal U}\) [GeV]  & $2.16\cdot 10^{16}$ & 
$0.02 \cdot 10^{16}$ \\
$M_{\nu_{R 3}}$ [GeV]           & $7.2 \cdot 10^{14}$ &
$[4.8,11]\cdot 10^{14}$ \\
\(m_{\nu_1}\) [eV]              & $3.5 \cdot 10^{-3}$ &
$[1.6,6.7]\cdot 10^{-3}$ \\
\hline
\end{tabular}
\caption{Reconstruction of SO(10), SUSY and neutrino parameters in an example scenario point (\(\mu>0\)).}
\label{tab:ParameterDetermination}
\end{table}
The masses of the staus are shifted relative to the masses of the first two generations by two terms 
\cite{Deppisch:2007xu,GUTreconstruction2,FPZ}:
\begin{eqnarray}
	m_{\tilde \tau_R}^2 - m_{\tilde e_R}^2  &=& 
	m_\tau^2 - 2\Delta_\tau,                            \\
	m_{\tilde \tau_L}^2 -  m_{\tilde e_L}^2 &=& 
	m_\tau^2 - \,\,\,\,\Delta_\tau - \Delta_{\nu_\tau}.
  \label{eq:RGthirdgen}
\end{eqnarray}
The shifts $\Delta_\tau$ and $\Delta_{\nu_\tau}$, generated by
loops involving charged lepton and neutrino superfields,
respectively, are given to leading order by
\begin{eqnarray}
	\Delta_\tau &\approx&
	\frac{m_\tau^2}{8\pi^2 v_d^2}
	\left(3M_0^2+A_0^2\right)
	\log\frac{\Lambda_{\mathcal U}^2}{\tilde{M}^2},
	\label{eq:DeltaTau}\\
	\Delta_{\nu_\tau} &\approx&
	\frac{m_t^2}{8\pi^2 v_u^2}
	\left(3M_0^2+A_0^2\right)
	\log\frac{\Lambda_{\mathcal U}^2}{M_{\nu_{R3}}^2},
  \label{eq:DeltaNuTau}
\end{eqnarray}
with the universal trilinear coupling \(A_0\) defined at the GUT scale.

Anticipating high-precision measurements at future colliders, such an SO(10) model can be investigated in central facets. As a concrete example, we study a scenario with SUSY parameters close to SPS1a/a$'$ \cite{spa,SPS} defined in Table~\ref{tab:ParameterDetermination}, compatible with the low-energy and cosmological data.

\begin{figure}
\begin{minipage}{\columnwidth}
\centering
\includegraphics[clip,width=0.8\textwidth]{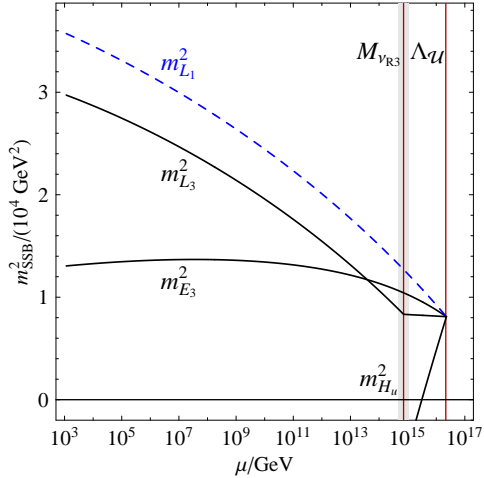}
\end{minipage}
\caption{Evolution of the first and third generation L,R slepton and Higgs mass parameters [\(D_{\mathcal{U}}=0\)].}
\label{fig:ScalarMassEvolutionGen1}
\end{figure}

The measurement of the slepton masses of the first two generations allows us to extract the common sfermion parameter $m_{16} = M_0$ as well as the D-term $D_{\mathcal{U}}$, {\it cf.} Eqs.~(6/7). Including the complete one-loop and the leading two-loop corrections, the evolution of the scalar mass parameters is displayed in Fig.~\ref{fig:ScalarMassEvolutionGen1} with the D-term set to zero for illustration. By extrapolating the measurements of slepton, Higgs and gaugino masses with the estimated experimental errors from the results of Refs.~\cite{spa,FPZ,a}, the high-scale parameters can be reconstructed with the precision shown in Tab.~\ref{tab:ParameterDetermination}. The RG evolution equations are evaluated to 2-loop order by means of the SPheno program \cite{Spheno}. The results indicate that the high-scale parameters $M_0$ and $D_{\mathcal{U}}$, driven by the slepton analysis, can be reconstructed at per-mill to per-cent accuracy.

The right-handed neutrino affects the logarithmic evolution of the mass parameter $m^2_{L_3}$ of the third generation in the most direct form. The characteristic difference in the evolution between $m^2_{L_3}$ and $m^2_{L_1}$ is exemplified in Fig.~\ref{fig:ScalarMassEvolutionGen1}. The position of the kink at $M_{\nu_{R3}}$ can be derived from the intersection of the parameter $\Delta_{\nu_\tau}$ as a function of $M_{\nu_{R3}}$, Eq.~\eqref{eq:DeltaNuTau}, with its measured value extracted from the slepton masses. Based on this estimate of $M_{\nu R_3}$, the value of the lightest neutrino mass is then determined via the seesaw mechanism, cf. Fig.~\ref{fig:Mivsm1}. The reconstructed values of $M_{\nu R_3}$ and $m_{\nu 1}$ in the SUSY scenario defined above, are shown in Tab.~\ref{tab:ParameterDetermination}, along with their uncertainties from propagating expected experimental errors. 

\subsection{Conclusion}
If the roots of physics are located near the Planck scale, experimental methods must be devised to explore the high-scale physics scenario including the grand unification of the Standard Model interactions. In this report, we have demonstrated how this goal can be achieved in a simplified SO(10) model with direct breaking to the SM gauge group. As naturally expected, the analysis of more complicated models will require a larger set of assumptions before the analysis can be performed, constrained however by additional observables like charged lepton masses, etc. Such an extended scheme is exemplified in a two-step breaking scenario SO(10) $\to$ SU(5) $\to$ SM analyzed in Ref.~\cite{Deppisch:2007xu}. The example described in this report has proved nevertheless that renormalization-group extrapolations based on high-precision results expected from Terascale experiments can provide essential elements for the reconstruction of the physics scenario near the GUT scale.

\end{document}